\documentclass[aps,prl,showpacs,twocolumn,floatfix,superscriptaddress]{revtex4}

\usepackage{amsmath} \usepackage{amsfonts} \usepackage{amssymb}
\usepackage{graphicx}
\usepackage{dcolumn}
\usepackage{bm}
\usepackage{color}
\usepackage[normalem]{ulem}

\begin{document}

\title{
%
%
Nonlinear localized modes in two-dimensional 
electrical lattices}

\author{L.Q. English}
\affiliation{%
Department of Physics and Astronomy,
Dickinson College, Carlisle, Pennsylvania, 17013  USA}

\author{F. Palmero}
\affiliation{%
Nonlinear Physics Group.
Escuela T\'{e}cnica Superior de Ingenier\'{\i}a Inform\'{a}tica.
Departamento de F\'{\i}sica Aplicada I.
Universidad de Sevilla.
Avda.~Reina Mercedes, s/n. 41012-Sevilla (Spain)
}

\author{J.F. Stormes}
\affiliation{Department of Physics and Astronomy, Dickinson College, Carlisle, Pennsylvania, 17013  USA}

\author{J. Cuevas}
\affiliation{%
Nonlinear Physics Group.
Escuela T\'{e}cnica Superior de Ingenier\'{\i}a Inform\'{a}tica.
Departamento de F\'{\i}sica Aplicada I.
Universidad de Sevilla.
Avda.~Reina Mercedes, s/n. 41012-Sevilla (Spain)
}

\author{R. Carretero-Gonz\'alez}
\affiliation{Nonlinear Dynamical Systems Group,
Department of Mathematics and Statistics, and Computational Science
Research Center, San Diego State University, San Diego CA, 92182-7720, USA}

\author{P.G. Kevrekidis}
\affiliation{Department of Mathematics and Statistics, University
of Massachusetts, Amherst, Massachusetts 01003-4515, USA}

\begin{abstract}
We report the observation of spontaneous localization of energy in two spatial dimensions in the context of nonlinear electrical lattices. Both {\it 
stationary and traveling} self-localized modes were generated experimentally and theoretically
in a family of two-dimensional {\it square}, as well as
{\it honeycomb} lattices composed of 6 $\times$ 6 elements. 
Specifically, we find regions in driver voltage and frequency where stationary discrete breathers, also known as intrinsic localized modes (ILM), exist and are stable due to the interplay of damping and spatially homogeneous driving. By introducing additional capacitors into the unit cell, these lattices 
can {\it controllably} induce traveling discrete breathers. When more than one such ILMs are experimentally generated in the lattice, the interplay of
nonlinearity, discreteness and wave interactions generate a complex dynamics 
wherein
the ILMs attempt to maintain a minimum distance between one another. 
%
%
Numerical simulations show good agreement with experimental results, and 
confirm that these phenomena qualitatively carry over to larger lattice sizes.  
\end{abstract}


\maketitle

{\it Introduction.} 
It has long been known that solitons emerge as classes of solutions to many nonlinear (lattice and partial) differential equations described chiefly by one spatial dimension; prominent examples are the KdV equation, the sine-Gordon or the nonlinear Schr\"{o}dinger equations. In two dimensions, quasi one-dimensional localization patterns can often still occur \cite{ablowitz}, but robust 
two-dimensional (2D) localization in continuous media is rather atypical 
(see Ref.~\cite{staliunas} and references therein). However, it is well-known that discreteness of the underlying medium can help to stabilize such localized solutions 
even in higher dimensions~\cite{flach,book}. Alternatively, one can 
externally enforce or introduce a periodicity in the form of a regular 
modulation in some property of the continuous (e.g.~in optical 
photorefractive or atomic Bose-Einstein condensate) media~\cite{moti,markus}, 
thus again breaking continuous translational symmetry. 
%

Here we show that two-dimensional 
discrete breathers, also known as  intrinsic localized modes (ILMs), 
{\it experimentally} exist and 
are stable in the context of two-dimensional, damped-driven electrical 
lattices. We characterize these breather and multi-breather states 
in parameter space, and we compare to numerical simulations and 
stability analysis. Finally, we focus particularly on versions of 
these lattices that support moving breathers. Discrete breathers
have been considered in a variety of other settings experimentally 
including (but not limited to)
micromechanical cantilever arrays~\cite{sievers}, 
Josephson-junction ladders (JJLs) \cite{alex},
granular crystals of beads interacting through Hertzian contacts~\cite{theo10}, layered antiferromagnetic crystals~\cite{lars3},
halide-bridged transition metal complexes~\cite{swanson},
 and dynamical models of the DNA double strand \cite{Peybi}. Yet,
in most of these examples, the coherent structures are effectively one-dimensional.
Even when higher dimensional (as is e.g.~possible in optical waveguide
arrays or photorefractive crystals~\cite{moti}), the states 
are typically stationary.
Hence, the experimental ability to systematically
generate discrete breathers in 
two-dimensional electrical lattices and, perhaps especially, to control their
mobility launches a new dimension in the modeling, and the theoretical
and experimental understanding of such states. This may be of broader
interest in other areas as well, such as 
JJLs (see e.g.~the theoretical proposal of Ref.~\cite{trias2}) and 
two-dimensional granular crystals (see e.g.~the recent experimental
realization of Ref.~\cite{daraio}). Another attractive feature of our
lattices is their potential square or honeycomb geometry, especially
since the latter has been a point of intense investigation at
both the linear and the nonlinear
level. This is due to its conical diffraction and Dirac (diabolical)
points examined intensely
in both the physical~\cite{mark2,moti2} and mathematical~\cite{miw}
communities, as well as due to potential connections with 
graphene nano-ribbons; see e.g.~\cite{kiv1,kiv2} and references therein.

\begin{figure}
\begin{center}
\includegraphics[width=2.5in]{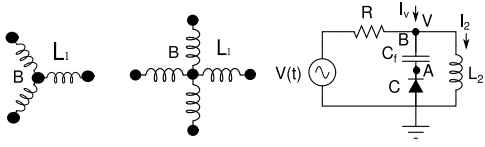}
\caption{Left: Schematic circuit diagrams of the basic geometry of the two electrical transmission lines (honeycomb and square), where black points represent single elements (right), with or without the block capacitor $C_f$.}
\label{line}
\end{center}
\end{figure}

{\it Experimental and theoretical setup.} The experimental system under investigation is a class of two-dimensional electrical lattices of either honeycomb or square geometry, as shown in Fig \ref{line}. These electrical lattices can be considered as a set of single cells representing nonlinear LC oscillators, each composed of a varactor diode (NTE 618) with a nonlinear capacitance $C(V)$ and an inductor $L_2= 330$ $\mu$H. These single cells are then coupled at point $V$ by inductors $L_1=680$ $\mu$H and are driven by a single sinusoidal voltage source $V(t)$ via a resistor $R=10$ k$\Omega$, with amplitude $V_d$ and frequency $f$. We study two different unit cell versions, with the only difference being the presence of a block capacitor between the diode and the coupling inductor in one of them, as explained in Ref.~\cite{eILM2}. The effect of the block capacitor is to make ILMs mobile; here we study two different values of block capacitances, $C_f=1$ $\mu$F and $C_f=15$ nF. The experimental lattices were 
comprised of 
36 elements with periodic boundary conditions, and node voltages $V_{n,m}$ were measured at a rate of 2.5 MHz using a multichannel analog-to-digital converter. In the linear limit of small-amplitude plane-waves,
%
we have obtained the dispersion relation, which yields as the lowest frequency (uniform mode) $f_{\rm min} \cong 312$ kHz, and the highest frequency  $f_{\rm max} \cong 689$ kHz in the square geometry configuration, and $f_{\rm max} \cong 617$ kHz in the honeycomb lattice. Notice that this highest-frequency mode in the linear spectrum is now above the second-harmonic of the uniform mode in square geometry lattice, and below it, but very close, in the honeycomb configuration.



Using basic circuit theory, in the block capacitor case, the dynamics 
of the lattice can be 
described by the equations \cite{Pal11},

\begin{eqnarray}
 \frac{d i_{n,m}}{d \tau}& = & \frac{L_2}{L_1} \left(\sum_{j,k} v^B_{j,k}-q\, v^B_{n,m}\right)-v^B_{n,m}, \nonumber \\
  \frac{d v^A_{n,m}}{d \tau}& = & \frac{1}{c(v^A_{n,m})}\bigg[i_{n,m}-i^D(v^A_{n,m})-\frac{v^B_{n,m}}{C_0\omega_0R_l}+ \nonumber \\ & & \frac{\cos(\Omega \tau)}
 {C_0\omega_0R}-\frac{v^A_{n,m}}{C_0\omega_0R} \bigg],\nonumber \nonumber \\
\frac{d v^B_{n,m}}{d \tau}& =&  \frac{d v^A_{n,m}}{d \tau}+\frac{C_0}{C_f}\left[i_{n,m}-\frac{v^B_{n,m}}{C_0\omega_0R_l}\right],
\label{model}
\end{eqnarray}



\noindent where the sum is taken over first, $q$, neighbors,  $q=4$ in square lattice case, and $q=3$ in honeycomb one. Also, the following dimensionless variables were used: $\tau=\omega_0 t$; $i_{n,m}=(I_v-I_2)/(\omega_0C_0V_d)$, where $I_v$ is the full current through the unit cell,  and $I_2$ the  current through the inductor $L_2$, both corresponding to cell $(n,m)$; $v=V/V_d$, used with superscripts A and B;
$V^B_{n,m}$ is the measured voltage at node $(n,m)$; $V^A_{n,m}$ is 
the voltage at an intermediate point between the varactor and the block 
capacitor; $\Omega=2 \pi f/\omega_0$,  $\omega_0=1/\sqrt{L_2C_0}$; $i_D=I_D/(\omega_0C_0V_d)$, where $I_D$ is the current through the varactor diode;  $c=C(V)/C_0$, where $C_0=C(0)$, and $C(V)$ is the nonlinear capacitance of the diode. Also, a phenomenological dissipation resistor, $R_l$, was included in the model to better approximate the experimental dynamics, and its value has been determined by matching the numerical and experimental nonlinear resonance curves corresponding to a single element \cite{Pal11}.  A simplified set of equations corresponding to the non--block capacitor case can be obtained by considering the limit $C_f\rightarrow \infty$. In all cases, the ratio $L_2/L_1$ 
characterizes the strength of the ``effective'' discreteness of the system.

\begin{figure}
\begin{center}
\includegraphics[width=4.0cm]{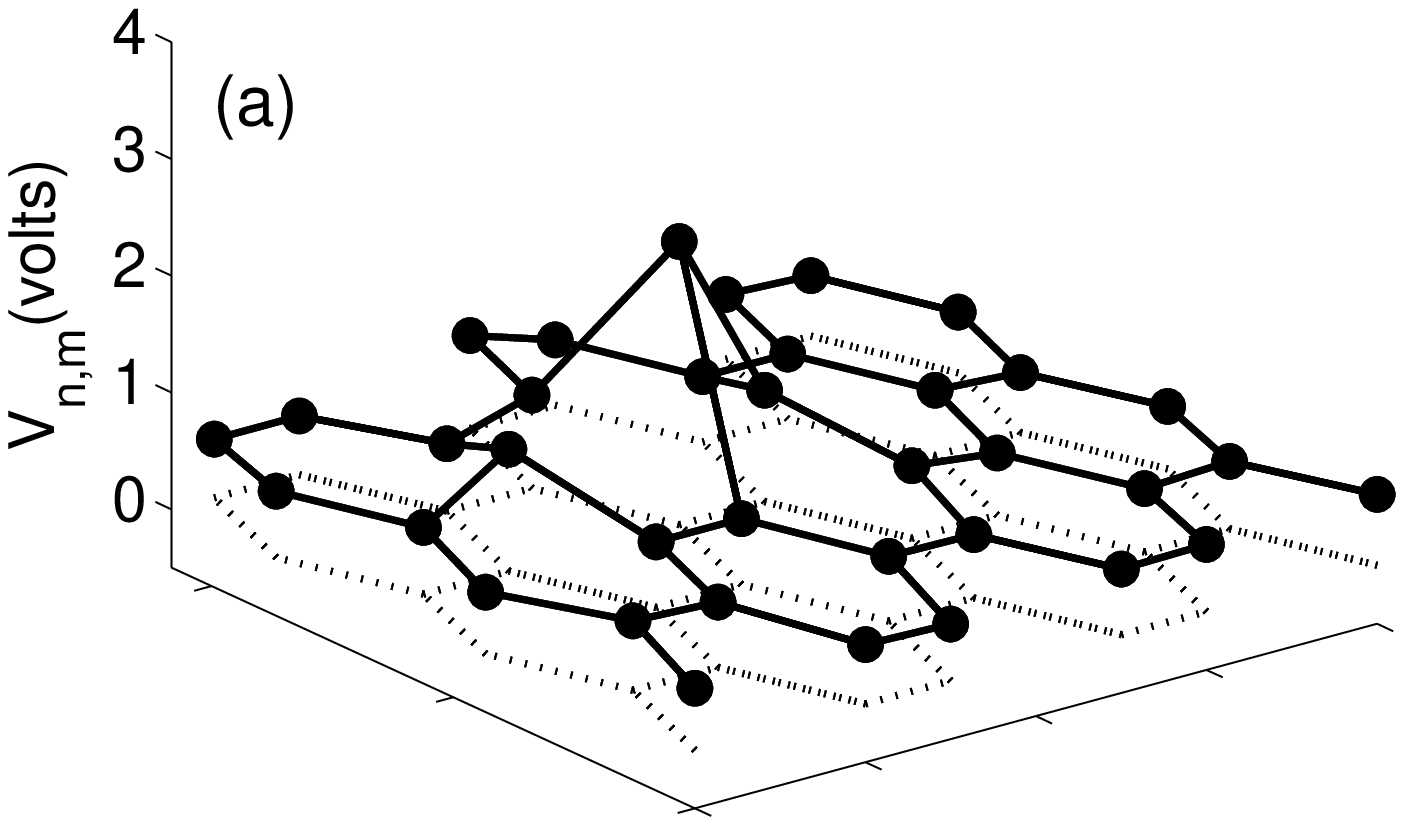}
\includegraphics[width=4.0cm]{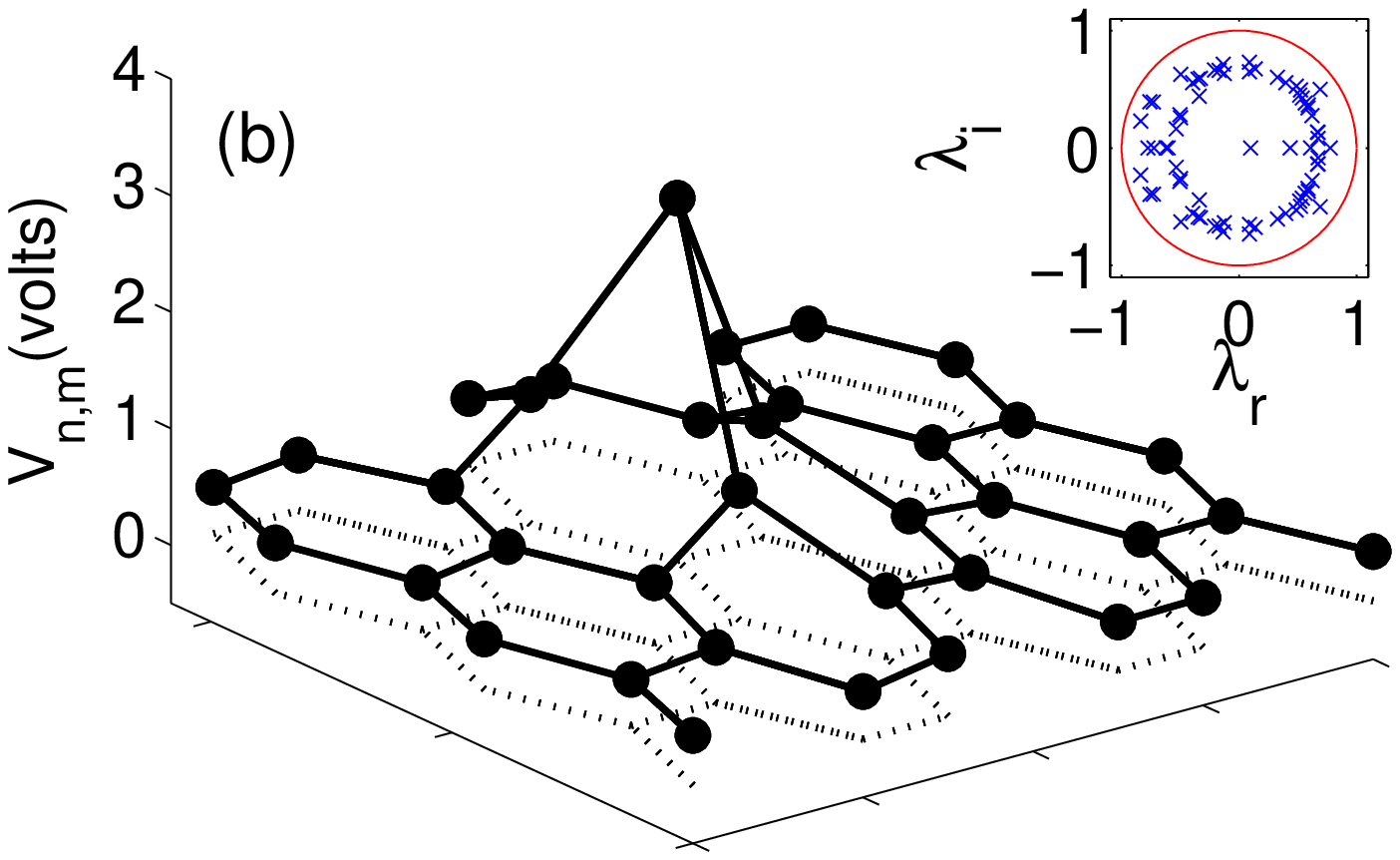}
\\
\includegraphics[width=4.0cm]{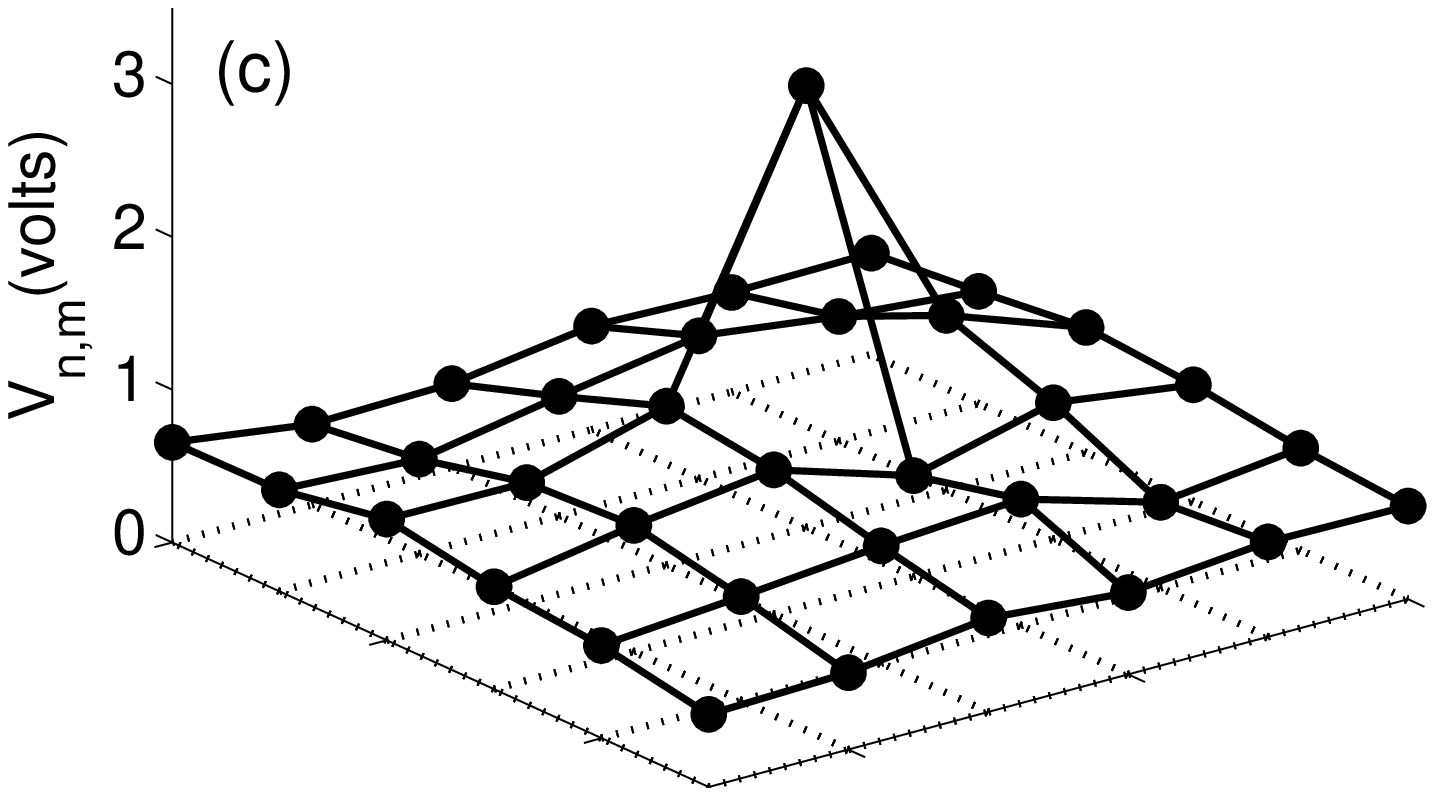}
\includegraphics[width=4.0cm]{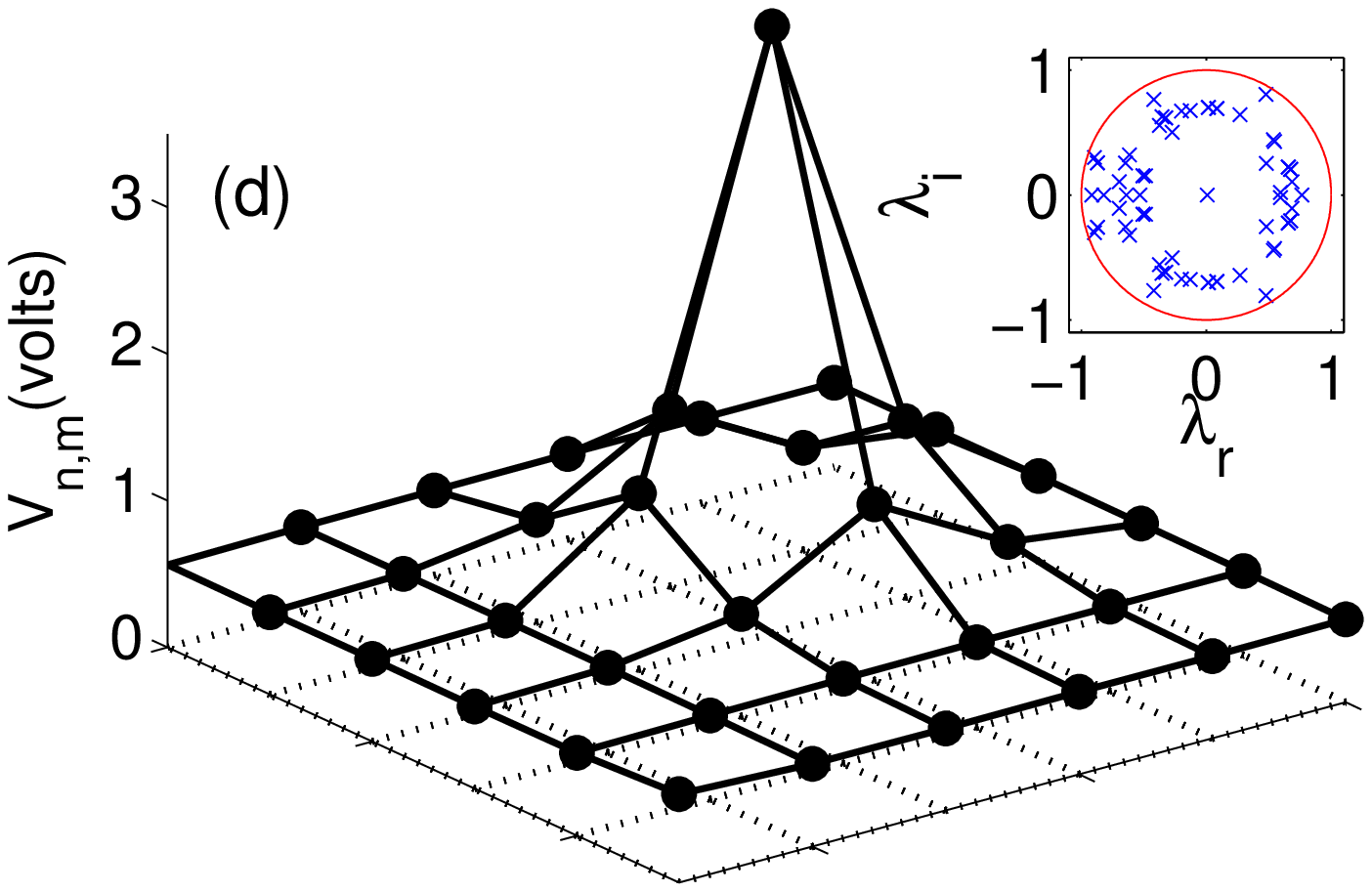}
\caption{(color online) 
Comparison between the experimental ((a) and (c)) and numerical 
((b) and (d)) profiles 
of stationary one peak breathers in a honeycomb lattice, (a) and (b),
and a square lattice, (c) and (d).
The insets show the Floquet numerical linearization spectrum, 
confirming (since all multipliers are inside the unit circle) 
the stability of solutions. The driver has amplitude and
frequency $V_d=1.5$ V and $f=283$ kHz in the honeycomb lattice case,
and $V_d=2$ V and $f=290$ kHz (experimental) and $f=301$ kHz (numerical) 
in the square lattice case.
}
\label{stationary}
\end{center}
\end{figure}

{\it Results on standing/traveling discrete breathers in square/honeycomb
2D lattices.} 
To generate discrete breathers, we have employed the 
well-known modulational instability of the driven uniform 
mode~\cite{flach}. In the case without block-capacitors, we obtain stationary and stable one-peak breathers, as shown in Fig.~\ref{stationary}. These 
localized modes are {\it robust}, persisting 
as long as the driver remains on. 

Numerics are generally found to be in
good agreement with experimental results. 
Yet, it can be seen in Fig.~\ref{stationary} that the amplitude of the one-peak breather is slightly higher in the numerics than in the experimental data. We attribute this slight mismatch to energy dissipation in the coupling inductors; these are stronger in the square lattice setting than in the honeycomb 
geometry, and they are not taken into account in our simplified model of Eq.~(\ref{model}). When this dissipation is included in our model, by means of some phenomenological small resistances in series with inductors $L_1$, 
this further improves the agreement for the ILM amplitude. For even higher amplitudes in the experiment, multi-peak breathers have been observed, which
can also be captured by the model.

\begin{figure}
\begin{center}
\includegraphics[height=2.1cm]{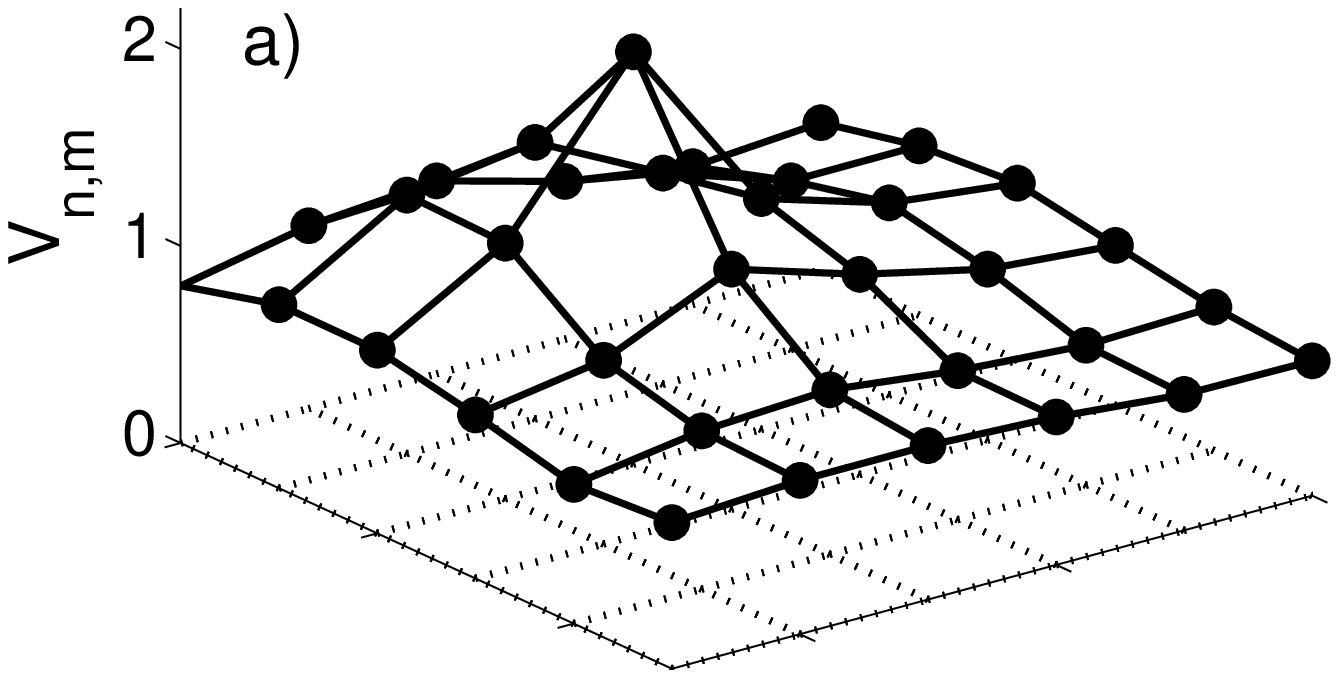}
~
\includegraphics[height=2.1cm]{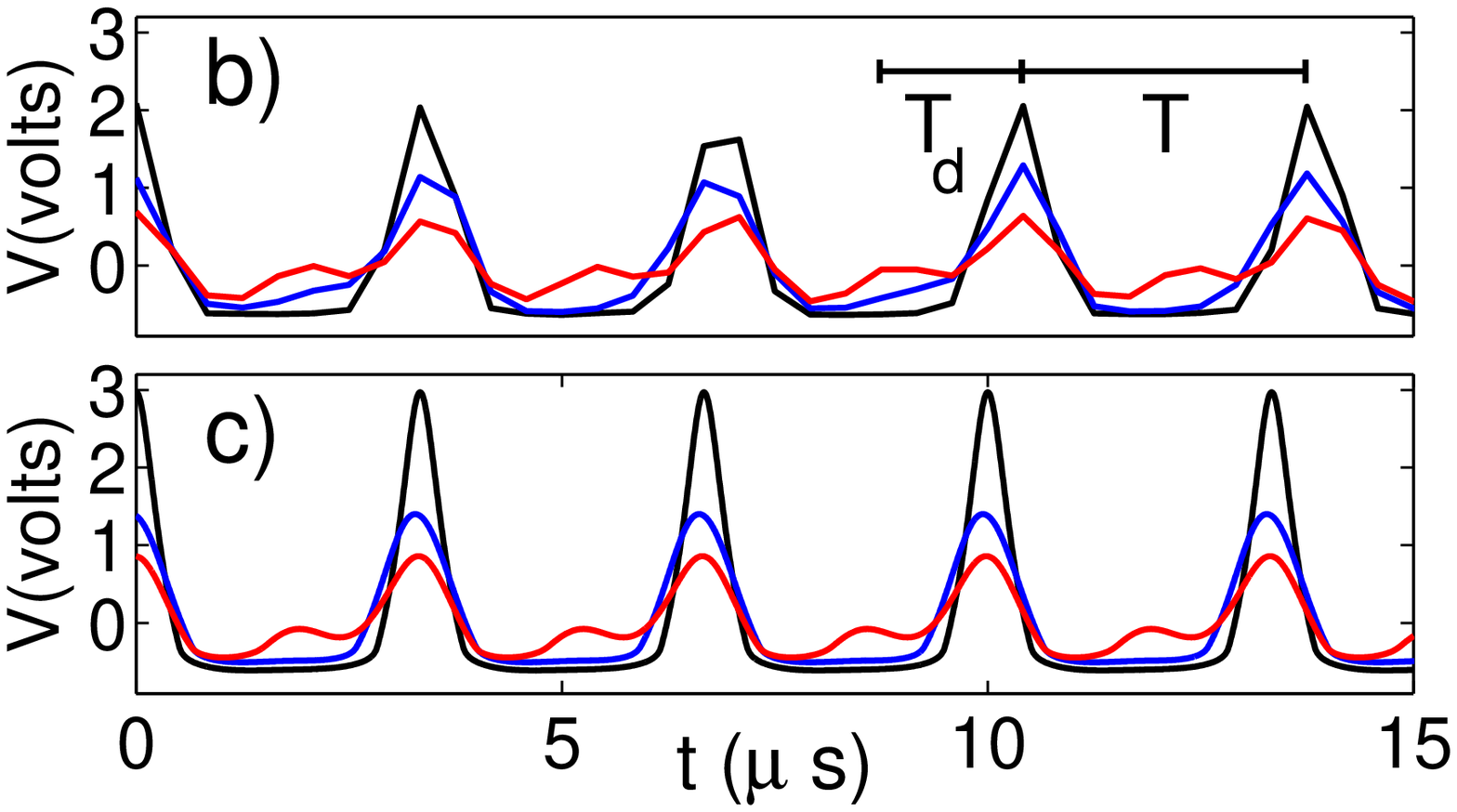}
\caption{(color online) Subharmonic breather corresponding to a square 
lattice, $V_d=6.4$ V and $f=600$ kHz. (a) Experimental spatial profile, 
and  (b) experimental and (c) numerical temporal oscillations of three 
different nodes: the ILM center (black), the first neighbor (blue), and 
the node furthest away from the ILM center (red). Note that
the oscillations of the node furthest away from the ILM center
display local maxima corresponding to the natural period $T$
of the ILM at twice the driving period $T_d = 1/f$.}
\label{sub}
\end{center}
\end{figure}

As in the one-dimensional line \cite{eILM3}, 
we again observe subharmonic breathers (experimentally and theoretically) 
in two dimensions. In Fig.~\ref{sub}, the peak of the breather oscillates with a frequency $f_{\rm ILM}=f/2$. Notably, in two dimensions it is generally more difficult to stabilize breathers via subharmonic driving ---the driving conditions have to fall into a fairly limited region of parameter-space. 
Further illustrating the fragility of subharmonic breathers in two dimensions, we note that the introduction of block capacitor destroys subharmonic breathers ---a qualitative difference with the one-dimensional lattice for which mobile subharmonic breathers were reported. The fact that the frequency of the subharmonic driver, namely twice the ILM response frequency, is now positioned within, or very close to, the linear dispersion band of the 2D lattice may explain this fragility.
Our numerical simulations indicate (results not shown here) that in 
larger lattices, multi-peak subharmonic breathers exist, and such subharmonic breathers appear to be more robust 
in the absence of block capacitors.

\begin{figure}
\begin{center}
\includegraphics[height=2.5cm]{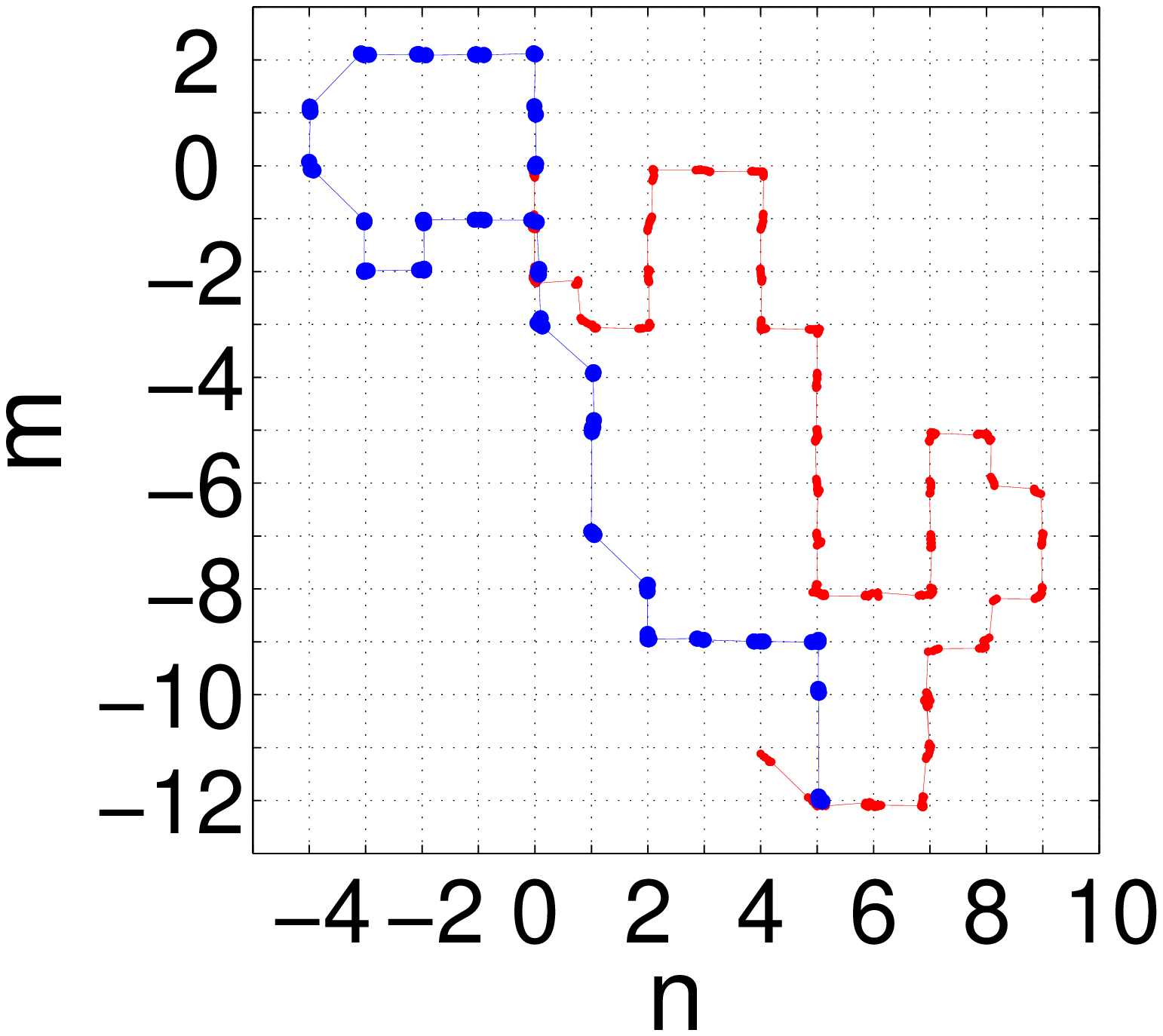}
~~~
\includegraphics[height=2.5cm]{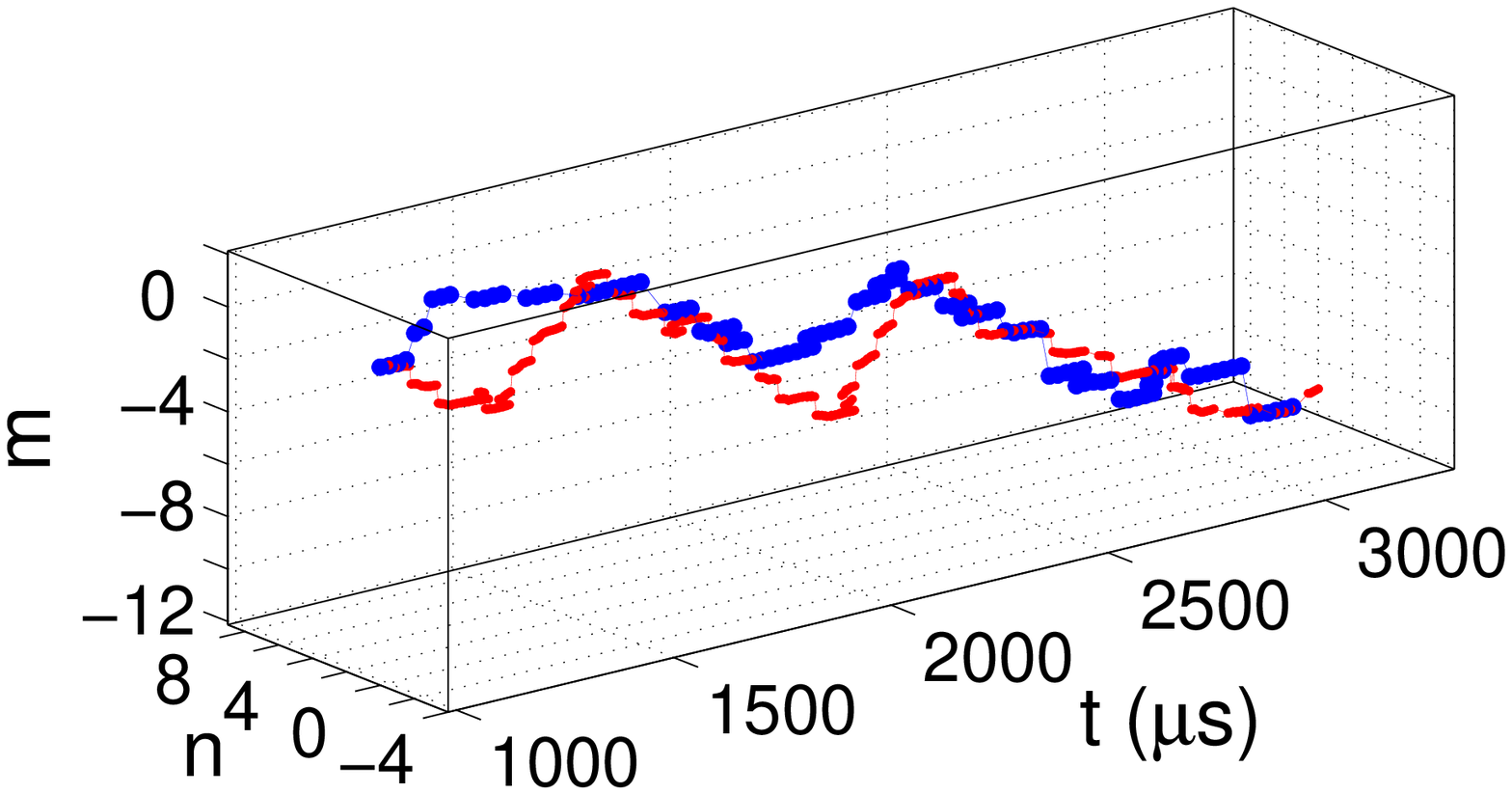}
\caption{(color online) Trajectory of a one peak moving breather 
on a square lattice for $V_d=2$ V, $f=300$ kHz and $C_f=1$ $\mu$F. 
Depicted are the positions of the center of mass of the breather
in the $(n,m)$ plane (left) and $(n,m,t)$ space (right).
The small (red) dots depict the experimental results
and the large (blue) dots depict the corresponding numerical simulations.
The trajectories have been unwrapped from the periodic lattice.
Solid lines have been added for guidance.
}
\label{mm}
\end{center}
\end{figure}

When a block capacitor of 1 $\mu$F is placed in series with the diode 
in the honeycomb and square lattice, the static breathers become mobile, as was
the case for  the one-dimensional chain \cite{eILM2}. 
Nevertheless, and contrary to the simpler situation in one dimension, 
where a clear direction of movement arises, the motion through the 
lattice appears to be more complex, as shown in Fig.~\ref{mm}. 
In general, numerical simulations obtained solving Eq.~(\ref{model}) 
show fairly good agreement with experiments and yield information 
about the intrinsic nature of this complex motion, where a strong 
sensitivity to small inhomogeneities is seen. The transition from 
one node to the next is such that 
where after a number of periods a noticeable asymmetry develops in the ILM 
profile; this leads to two neighboring nodes attaining equal 
amplitude, and finally the ILM becoming centered (initially between
two sites and then) on the next node. This sequence is now also 
observed in two dimensions. 
%
Notice that the mobility problem has been argued to be
quite important in other non-square lattices, such as the hexagonal
one where breather mobility was proposed as responsible for the presence of
dark lines in natural crystals of muscovite mica~\cite{MarinEilbeckRussell:98}, and
reconstructive transformations in layered silicates~\cite{Arc06}. 
In our square electrical lattice, we generally observe (cf.~Fig.~\ref{mm})
a directed motion which is
interrupted by (longer) intervals of localization as the wave 
struggles to overcome the well-known Peierls-Nabarro (PN) barrier~\cite{flach}.
Note that given the small size of the lattice, in addition
to the role of inhomogeneities, small amplitude residual 
excitations (``phonons'') are also important in directing the breather
motion.
It is also worth mentioning that, although the movement along 
the square lattice usually happens along the edges of the lattice,
some transitions are also observed to happen, both
in the experiment and the numerics, along the
diagonals that have a slightly higher PN barrier.
Despite the complex nature of the motion of the breathers,
the organizing principle obeyed in both one and two dimensions 
is that the ILM never hops back to the node it occupied prior to its 
current location due to some residual capacitor-charge impurity associated with that 
node. In one dimension, this principle necessarily gives rise to 
orderly, uni-directional motion. 
In two-dimensional hexagonal 
lattices there are still two choices available to the ILM, and 
in a square lattice there are three. Thus, the motion does not 
have to be uni-directional.
It would be interesting to statistically measure the ``diffusion''
associated with the apparent irregular trajectories displayed
by the moving breathers. Although the system is deterministic,
due to small defects and residual excitations, the breathers
seem to follow a biased erratic motion with the 
constraint of not going back in the direction they came from. 
Comparing the scaling of the diffusion associated with this 
peculiar behavior with the classical Brownian motion
and biased random walk variants thereof 
would be an interesting avenue for future exploration.

\begin{figure}
\begin{center}
\includegraphics[width=4cm]{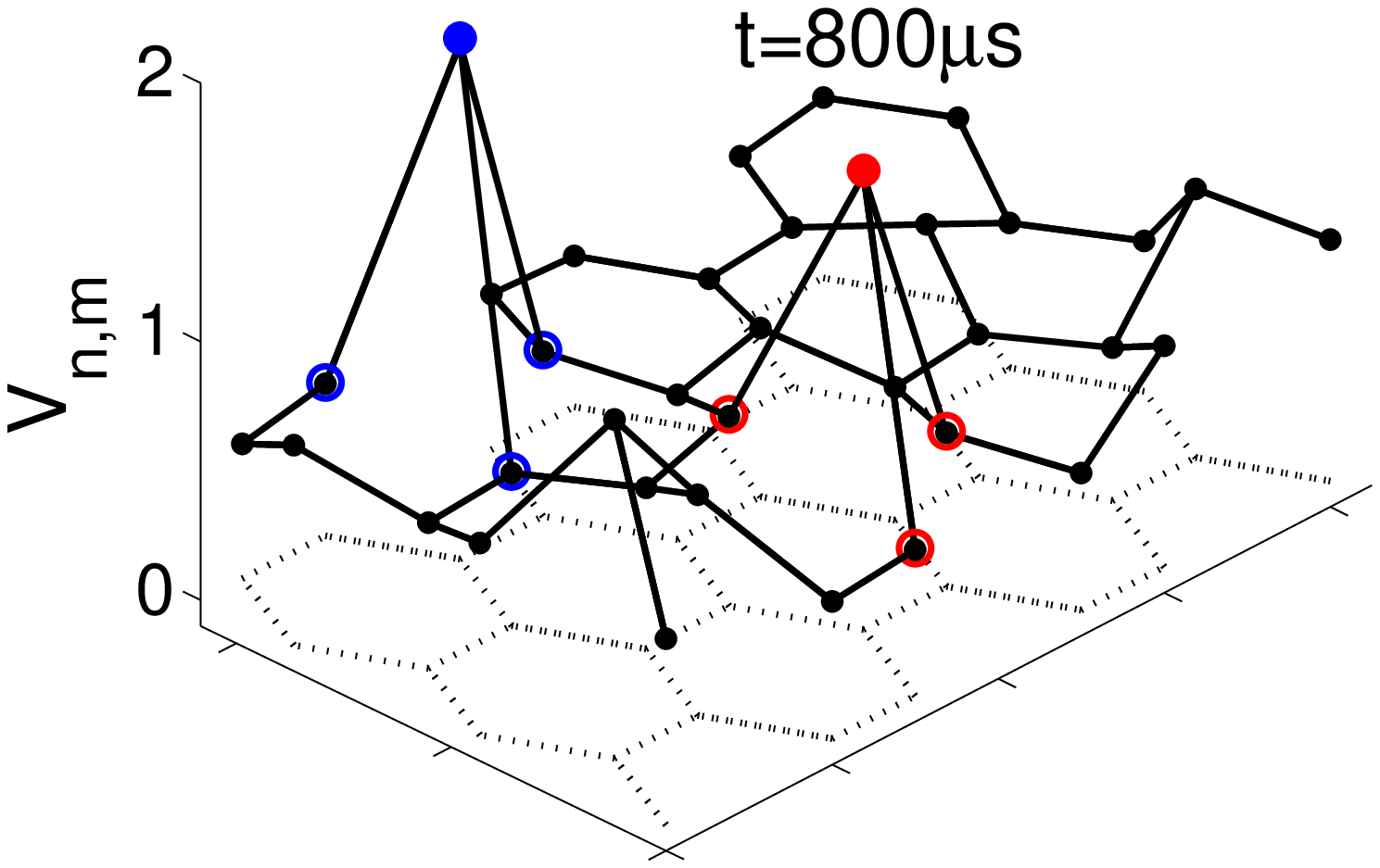}
~~~
\includegraphics[width=4cm]{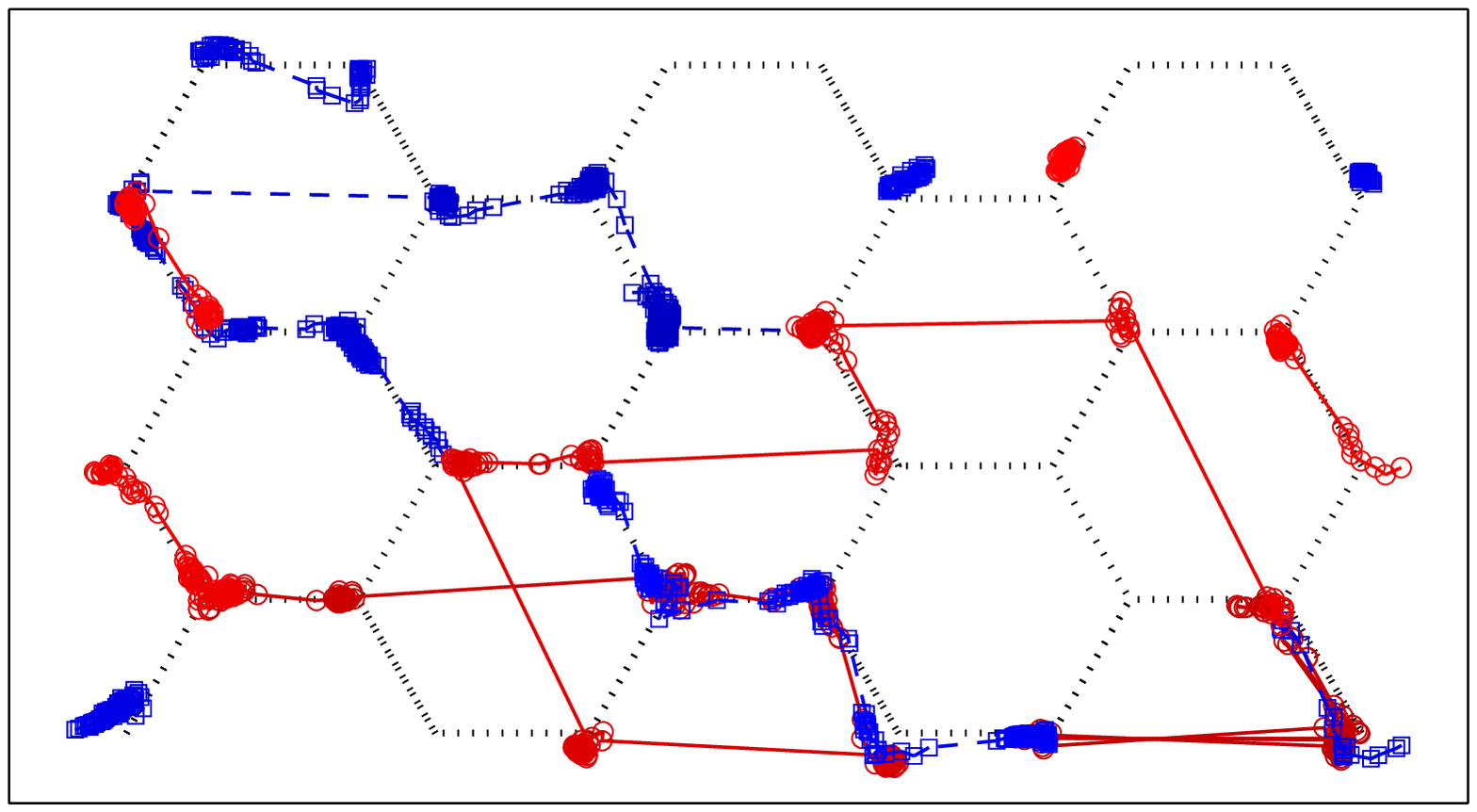}
\caption{(color online) 
Left:
Experimental two peak moving breather profile on a honeycomb lattice with
periodic boundary conditions for $V_d=2.5$ V, $f=330$ kHz, and  $C_f=1$ $\mu$F.
Right: Trajectory of the corresponding centers of mass of the two interacting
breathers (circles [red] and squares [blue]).
Solid lines have been added for guidance.
}
\label{mm2}
\end{center}
\end{figure}

In general, multi-peak breathers, or collections of breathers, 
also become mobile in the lattices with block capacitors. In all 
of our experimental results, the relevant motion is complex
yet the relative distance between peaks remains fairly constant 
and does not fall below a minimum value (two edges away)
(cf.~Fig.~\ref{mm2}). 
This observation suggests that two ILMs tend to repel 
each other upon close proximity.
We also note, as in the square lattice, that although most transitions
during motion happen through the lattice edges, there are some
transitions that happen along the long diagonal of the
honeycomb cell (cf.~Fig.~\ref{mm2}). We note that we did not observe 
transitions along the short diagonals of the honeycomb cell. 

\begin{figure}
\begin{center}
\includegraphics[height=1.7cm]{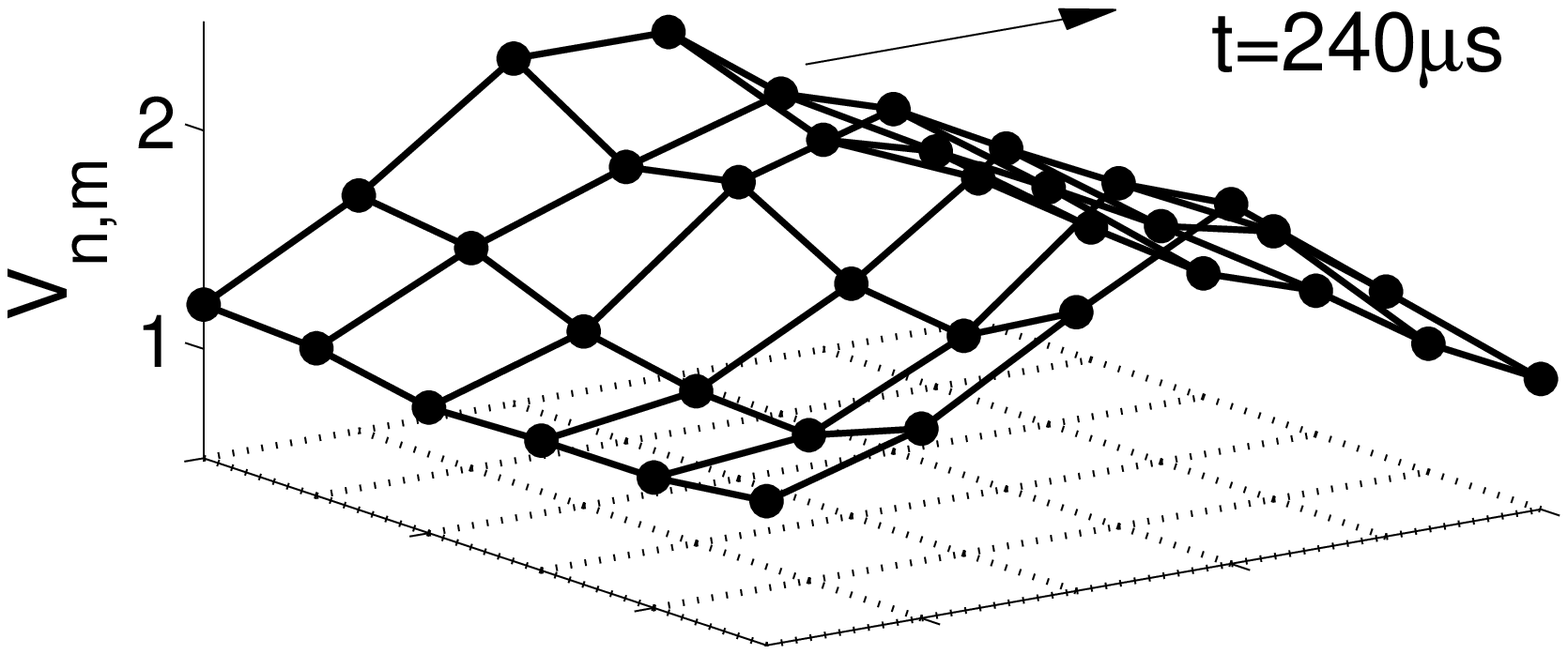}
\includegraphics[height=1.7cm]{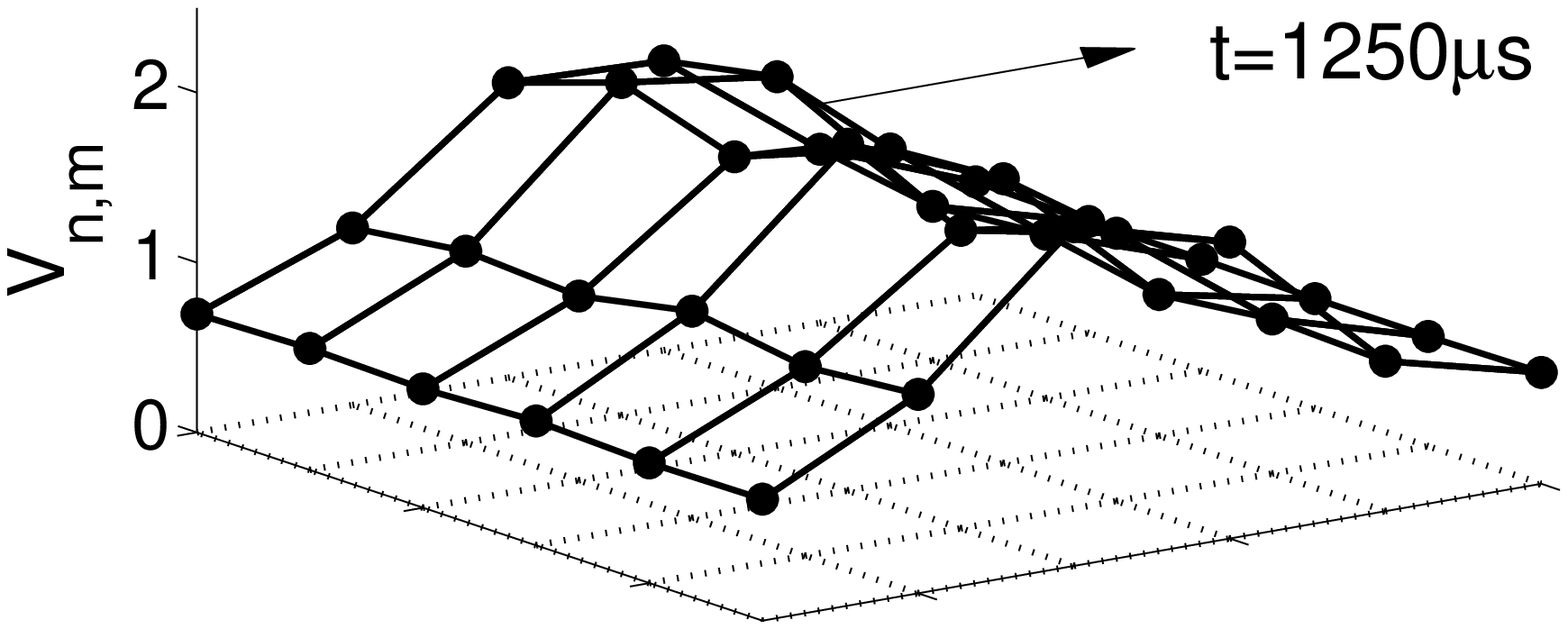}
\caption{
Experimental (left) and numerical (right) moving front profile for 
a $6\times 6$ square lattice.
$V_d=2$ V, $f=300$ kHz and $C_f=15$ nF.}
\label{front}
\end{center}
\end{figure}

A different scenario takes place when a smaller block capacitance, 
$C_f=15$ nF, is chosen. In that case, breathers become mobile but 
quickly collapse into a collective pattern resembling a planar 
(one-dimensional) wavefront, which rapidly transports energy 
coherently through the system, as shown in Fig.~\ref{front}. 
%
%
It is intriguing that a spatially homogeneous driver can sustain 
such nonlinear collective patterns characterized by energy transport. 
%
%
We should point out that the types of coherent patterns experimentally 
observable in the honeycomb lattice are apparently more complex than 
in the square lattice where only true planar wavefronts are seen. For 
the honeycomb lattice the equivalent planar waves, because of the 
geometry of the lattice, look more complex, and can be found
at particular driving frequencies and amplitudes (results not shown here).
%
%
Numerical simulations performed on larger square lattices (results not shown here) 
suggest that one-dimensional wavefront profiles are not robust
---presumably due to transverse instabilities--- and break down into
patterns consisting of several ILMs, each one extending over 
a number of sites. 
A more detailed account of the phenomenology in larger lattices
is currently under investigation and will be reported in a
future publication.
%

{\it Conclusions \& future challenges.} 
In summary, we have generated two-dimensional discrete breathers 
in the setting of  damped-driven electrical lattices. To our knowledge, 
this is the first time that breathers have been experimentally stabilized 
in higher-dimensional discrete lattices by direct (and subharmonic) 
driving, with motion systematically induced to them and that 
these features have been examined not only in square but importantly 
also in honeycomb lattices. 

We have characterized the statics, stability and mobility
of these modes of self-localized 
energy, centered over a particular node of the 
lattice, and extending over a few lattice sites.
%
The breathers have been found to persist indefinitely, and are either 
stationary or hopping in the lattice, depending on the precise make-up 
of the unit cell. 
%
%
We have observed that interacting breathers seem to  repel
each other when placed in close proximity which has the effect of 
(roughly) maintaining their pairwise distance in small periodic lattices.

This work paves the way for numerous studies. These include 
the detailed characterization of breathers, their stability and mobility 
properties and associated potential barrier in long (and infinite) 2D chains; 
the characterization of impurities and their role in inducing mobility; 
the examination of spectral gap and higher gap (nonlinear) states, 
especially in honeycomb lattices;
and
the generalization of such chains even in fully 3D configurations.
These issues of broader interest to other settings such as 
Josephson-junction ladders~\cite{trias2}
and granular crystals~\cite{daraio} will be considered in future publications.

This research was supported by the Ministerio de Ciencia e Innovaci\'on 
of Spain (FIS2008-04848), University of Seville (plan propio),  
the US-NSF via CMMI-1000337, and US-AFOSR via FA9550-12-1-0332.  
F.P.~acknowledges Dickinson College for hospitality.

\end{document}